# Crystal Structure and Thermoelectric Transport Properties of As-Doped Layered Pnictogen Oxyselenides NdO$_{0.8}$F$_{0.2}$Sb$_{1-x}$As$_x$Se$_2$

Kota Morino,[1] Yosuke Goto,[1]* Akira Miura,[2] Chikako Moriyoshi,[3] Yoshihiro Kuroiwa,[3] and Yoshikazu Mizuguchi[1]

[1]Department of Physics, Tokyo Metropolitan University, 1-1 Minami-osawa, Hachioji, Tokyo 192-0397, Japan

[2] Faculty of Engineering, Hokkaido University, Kita 13, Nishi 8 Sapporo 060-8628, Japan

[3] Department of Physical Science, Hiroshima University, 1-3-1 Kagamiyama, Higashihiroshima, Hiroshima 739-8526, Japan

We report the synthesis and thermoelectric transport properties of As-doped layered pnictogen oxyselenides NdO$_{0.8}$F$_{0.2}$Sb$_{1-x}$As$_x$Se$_2$ ($x \leq 0.6$), which are predicted to show high-performance thermoelectric properties based on first-principles calculation. The crystal structure of these compounds belongs to the tetragonal *P*4/*nmm* space group (No. 129) at room temperature. The lattice parameter *c* decreases with increasing *x*, while *a* remains almost unchanged among the samples. Despite isovalent substitution of As for Sb, electrical resistivity significantly rises with increasing *x*. Very low thermal conductivity of less than 0.8 Wm$^{-1}$K$^{-1}$ is observed at temperatures between 300 and 673 K for all the examined samples. For As-doped samples, the thermal conductivity further decreases above 600 K. Temperature-dependent synchrotron X-ray diffraction indicates that an anomaly also occurs in the *c*-axis length at around 600 K, which may relate to the thermal transport properties.

**Keywords:** thermoelectric materials; mixed-anion compounds; crystal structure analysis; transport properties

*Correspondence: y_goto@tmu.ac.jp

**Introduction**

A thermoelectric device is a solid-state device that can directly convert heat to electricity or vice versa without any gas or liquid working fluid [1–5]. The efficiency of the thermoelectric device is primarily governed by the material's dimensionless figure of merit, $ZT = S^2 T \rho^{-1} \kappa^{-1}$, where $T$ is the absolute temperature, $S$ is the Seebeck coefficient, $\rho$ is the electrical resistivity, and $\kappa$ is the thermal conductivity. Because these transport properties strongly correlate with each other, it is not easy to achieve high-$ZT$ thermoelectric materials. A promising approach to developing novel thermoelectric materials is by utilizing the first-principles calculation of transport properties under some assumptions, such as a constant relaxation time [6–13]. Indeed, several candidate materials have been theoretically predicted to show high thermoelectric performance, and some have been experimentally investigated. However, the verification of these theoretical predictions is insufficient because some of the promising materials cannot be obtained in a thermodynamically stable phase. Furthermore, even if the parent materials are obtained, tuning carrier density in order to optimize its thermoelectric performance may be hampered by issues including self-compensation due to intrinsic point defects [14–16]. It remains important to experimentally investigate the new material systems to develop high-performance thermoelectric materials.

The layered pnictogen oxychalcogenides REOPnCh$_2$ (RE = rare earth, Pn = pnictogen, Ch = chalcogen) are one of the materials predicted to show enhanced thermoelectric properties [17,18]. The crystal structure of REOPnCh$_2$ is characterized by alternate stacks of PnCh$_2$ conducting layers and REO spacer layers, illustrated schematically in Figure 1. These compounds have been extensively studied because of their superconductivity when Pn = Bi [19–21]. Moreover, the thermoelectric figure of merit $ZT$ reached 0.36 for LaOBiSSe [22]. Although the reported $ZT$ is still moderate, their striking features of (1) intrinsically low lattice thermal conductivity, 1–2 Wm$^{-1}$K$^{-1}$ at 300 K, and (2) a tolerance for various elemental substitution, offer promise for efficient thermoelectric materials [23–26]. In particular, low lattice thermal conductivity is achieved due to the rattling motion of Bi, even though these compounds have no oversized cage [27,28]. To enhance their thermoelectric power factor ($S^2 \rho^{-1}$), the guiding principles are indicated on the basis of first-principles calculation: (1) Small spin–orbit coupling, (2) small Pn–Pn and large Pn–Ch hopping amplitudes, and (3) a small onsite energy difference between the Pn–$p_{xy}$ and Ch–$p_{xy}$ orbitals [17,18]. Consequently, employing light Pn and heavy Ch is believed to result in an increase in thermoelectric performance. It has been predicted that REOAsSe$_2$, which is yet to be synthesized, shows $ZT$ exceeding 2 as a result of the above-mentioned optimization.

So far, the replacement of Pn with a light element has been demonstrated using Pn = Sb [29–34]. We have reported the synthesis and transport properties of RE(O,F)SbSe$_2$ for RE = La, Ce, and Nd [30,31]. Although aliovalent ion substitution of F at the O site implies electron doping in these compounds, electrical resistivity was very high when Pn = Sb. Indeed, electrical resistivity drastically decreased by Bi substitution at the Sb site, and a superconducting transition under ultra-high-pressure was demonstrated in NdO$_{0.8}$F$_{0.2}$Sb$_{1-x}$Bi$_x$Se$_2$ [34].

In this study, we report the synthesis of As-doped NdO$_{0.8}$F$_{0.2}$SbSe$_2$, namely, NdO$_{0.8}$F$_{0.2}$Sb$_{1-x}$As$_x$Se$_2$ ($x \leq 0.6$). The crystal structure of these compounds belongs to the tetragonal *P*4/*nmm* space group (No. 129). Electrical resistivity significantly increases as a result of As substitution, and very low thermal conductivity of less than 0.8 Wm$^{-1}$K$^{-1}$ is observed in all samples. We also discuss an anomaly observed in the temperature dependence of the lattice parameter $c$ and thermal conductivity at around 600 K.

**Experiments**

Polycrystalline samples were synthesized using solid-state reactions employing dehydrated Nd$_2$O$_3$, NdSe, NdSe$_2$, Sb (99.9%), As (99.999%), and Se (99.999%) as precursors. The dehydrated Nd$_2$O$_3$ was prepared by heating commercial Nd$_2$O$_3$ powder (99.9%) at 700 °C for 15 h in air. The mixture of NdSe and NdSe$_2$ was prepared by the reaction of Nd (99.9%) and Se in a molar ratio of 2:3, heated at 500 °C in an evacuated silica tube. A stoichiometric mixture of these precursors was pressed into a pellet and heated at 600−700 °C for 20 h. These procedures were conducted in an Ar-filled glovebox with a gas-purifier system. The samples were densified by hot pressing at 50 MPa at 600 °C for 30 min. The relative density of the obtained samples was calculated as >95%.

The chemical compositions of the samples were analyzed using an energy dispersive X-ray spectrometer (EDX; SwiftED3000, Oxford instruments, Abingdon-on-Thames, Oxfordshire, UK). The phase purity and crystal structure of the samples were examined using powder X-ray diffraction (XRD) with Cu K$\alpha$ radiation (Miniflex 600 equipped with a D/teX Ultra detector, Rigaku, Akishima, Tokyo, Japan). Synchrotron powder X-ray diffraction (SXRD) was measured at the BL02B2 beamline of SPring-8 (proposal number 2019B1195). The diffraction data were collected using a high-resolution one-dimensional semiconductor detector (multiple MYTHEN system, Dectris, Täfernweg, Baden, Switzerland) [35]. The wavelength of the radiation beam was determined to be 0.496391(1) Å using a $CeO_2$ standard. The temperature dependence of SXRD was collected using a nitrogen gas blower up to 694 K. The crystal structure parameters were determined by Rietveld analysis using the Rietveld refinement program RIETAN-FP [36]. The crystal structure was depicted using VESTA [37].

The electrical resistivity $\rho$ and Seebeck coefficient $S$ were examined using the four-probe method and quasi-steady-state method, respectively, under a helium atmosphere (ZEM-3, Advance Riko, Yokohama, Kanagawa, Japan). The samples used for the measurements were rectangular bars with a size of approximately 2 × 3 × 10 $mm^3$. The thermal conductivity $\kappa$ was obtained using the relationship $\kappa = DC_p d_s$, where $D$, $C_p$, and $d_s$ are the thermal diffusivity, specific heat, and sample density, respectively. The thermal diffusivity was measured by a laser flash method (TC1200-RH, Advance Riko). The samples used for the measurements were pellet-shaped disks with a diameter of ϕ10 mm and a thickness of 2 mm. The value of $C_p$ was estimated by the Dulong–Petit model, $C_p = 3nR$, where $n$ is the number of atoms per formula unit, and $R$ is the gas constant.

## Results and Discussion
### Sample Characterization and Crystal Structure Analysis

Figure 2 shows XRD patterns for $x$ = 0–0.6. Almost all the diffraction peaks are attributable to those of the tetragonal $P4/nmm$ space group (No. 129), indicating the samples are dominantly composed of $NdO_{0.8}F_{0.2}SbSe_2$-type phase. However, several diffraction peaks due to the impurity phases are also observed. Rietveld analysis shows that the amount of impurities of $x$ = 0.6 is as follows: Sb in the $P2_1/m$ space group (2.4 wt.%), $NdO_{0.67}F_{1.66}$ (2.0 wt.%), $Nd_4O_4Se_3$ (1.4 wt.%), and Sb in $R\bar{3}m$ space group (0.8 wt.%.) (Figure 3). Because the amount of impurities significantly increased for $x$ = 0.7, we employed $x \leq 0.6$ in further studies. Figure 4 shows the chemical composition obtained using EDX. With increasing $x$, the amount of analyzed Sb decreases, while that of As increases. The amount of Se remains almost unchanged among the samples. The EDX confirms that the chemical composition of the present samples is in reasonable accordance with the nominal composition of starting materials.

Figure 5 shows the lattice parameters $a$ and $c$ as a function of $x$. With increasing $x$, the

$c$ decreases almost monotonically, while $a$ increases only slightly. The decrease of $c$ is most likely due to the smaller ionic radius of As than that of Sb. For example, Shannon's ionic radius of $As^{3+}$ and $Sb^{3+}$ with six coordinates is as follows: $As^{3+}$ = 58 pm and $Sb^{3+}$ = 76 pm [38], although the valence state of As and Sb is still under investigation.

Figure 6 shows the selected bond distances and angles. The bond distance for in-plane Sb/As (Pn)-Se1 is almost independent of $x$, consistent with the almost unchanged lattice parameter $a$. The bond distance for Pn-Se2 decreases almost linearly with increasing $x$, which is likely related to a decrease of lattice parameter $c$. Bond distances for interplane Pn-Se1 tends to increase with an increase of $x$, but shows a slightly complex behavior. This is likely due to the contribution of the following opposite trends: (1) A decrease in $c$-axis length appears to reduce the interplane Pn-Se1 distance, and (2) the Se1-Pn-Se1 bond angle tends to approach 180° with increasing $x$, as shown in Figure 6d, which tends to increase the interplane Pn-Se1 distance.

*Thermoelectric Transport Properties*

The temperature dependence of electrical resistivity is depicted in Figure 7a. The resistivity is significantly increased as a result of As substitution, despite isovalent substitution of As into Sb. Although the origin of the insulating nature in these compounds is not yet clear, the results evoke the decrease of resistivity by Bi substitution into the Sb site in $NdO_{0.8}F_{0.2}Sb_{1-x}Bi_xSe_2$ [31]. A negative Seebeck coefficient indicates the *n*-type polarity of these samples, as shown in Figure 7b. The very high absolute value of the Seebeck coefficient is in agreement with the insulating behavior observed in electrical resistivity. The magnitude of the Seebeck coefficient decreases rapidly with increasing temperature, suggesting small amounts of minority carriers (holes) are also excited at high temperatures.

At first glance, the insulating nature observed in $NdO_{0.8}F_{0.2}Sb_{1-x}As_xSe_2$ seems to contradict first-principles calculation, which indicates metallic band structure [34]. One may deduce that this is due to insufficient electron doping through a spacer layer, because it is difficult to determine the O and F contents using EDX or XRD. However, even if the actual F content is lower than the nominal composition of starting materials, electron doping should trigger metallic conduction, as expected in a conventional semiconductor. In $BiCh_2$-based layered compounds, which are structural analogs of the present compounds, it is revealed that emerging bulk superconductivity requires the suppression of local disorder, as well as electron doping [3]. That is to say, some $BiCh_2$-based compounds do not show bulk superconductivity, although the electron is sufficiently doped by a spacer layer. For example, the extended X-ray absorption fine structure shows that mean square relative displacement (MSRD) of in-plane Bi-Ch bond distance is a measure of local disorder [40–42]. Namely, a large MSRD indicates the existence of disorder in the Bi-Ch layer, which is detrimental to inducing bulk superconductivity. Furthermore, pair distribution function analysis using neutron and synchrotron X-ray diffraction also indicates local symmetry lowering in the tetragonal structure [43–46]. Detailed crystal structure analysis using Rietveld refinement shows that this local disorder/distortion is also observed in terms of a large atomic displacement parameter [47]. Indeed, first-principles calculation indicates the tetragonal *P*4/*nmm* structure is not energetically favored, and shows the existence of imaginary phonons in this structure [48–51]. All of these analyses strongly suggest that the doped electron is not activated because of local disorder/distortion in these compounds.

To suppress the local disorder/distortion, it is believed that applying in-plane chemical pressure that promotes overlapping between Pn *p* and Ch *p* orbitals is useful by chemical doping [3,39]. We briefly describe our preliminary doping study results below to achieve more conductive samples. First, we examined Te doping into the Se site, but the solubility limit of Te was estimated at less than 10%. We then investigated Gd/Sm doping into the Nd site to shrink the lattice parameter *a* by using lanthanide contraction, and the solubility limit was approximately 30%. The decrease in electrical resistivity as a result of this doping was not significant, suggesting that there was still local disorder/distortion in these samples. Local structure analysis will help provide insights into the insulating properties of samples in the present study.

Figure 8 shows the temperature dependence of thermal conductivity. Very low thermal conductivity of less than 0.8 Wm$^{-1}$K$^{-1}$ is obtained for all samples. Because of its insulating nature, the electronic component in thermal conductivity is negligibly small. According to the Wiedemann–Franz relationship, electronic thermal conductivity is expressed as $\kappa_e = LT\rho^{-1}$, where *L* is the Lorenz number (typically $L = 2.44 \times 10^{-8}$ V$^{-2}$K$^{-2}$). The $\kappa_e$ of the present samples is estimated to be less than 1% of total thermal conductivity. Room-temperature thermal conductivity further decreases as a result of As substitution, likely because of point defect scattering of the phonon in solid solution [51–58]. The thermal conductivity decreases with increasing temperature, indicating that the phonon scattering process is dominated by the Umklapp process. Notably, the thermal conductivity of As-doped samples rapidly decreases above 600 K, as shown in the inset of Figure 8, suggesting a change in phonon properties at this temperature. Here, the measurement results on the heating and cooling run almost coincide.

**High-Temperature Synchrotron X-ray Diffraction**

To further study this anomaly at around 600 K, we performed temperature-dependent SXRD measurement for $x = 0.2$. Figure 9a shows SXRD patterns at temperatures between 350 and 694 K with a heating rate of 30 Kmin$^{-1}$. Diffraction peaks attributable to impurities are not significantly increased during the measurements. The *a*-axis length increases monotonically with temperature because of thermal expansion, as shown in Figure 9b. On the other hand, the *c*-axis length shows a kink at 600 K, at which point thermal conductivity also changes anomalously (Figure 9c and the inset of Figure 8). Although we measured SXRD in the heating run only, room temperature lattice parameters of the recovered sample are consistent with those before the measurements. This strongly suggests that the obtained results are intrinsic properties of this sample. Note that this result is somewhat similar to the phase transition from tetragonal to collapsed tetragonal phase, as demonstrated in 122-type compounds such as CeFe$_2$As$_2$ and LaRu$_2$P$_2$ [59,60]. However, a more detailed study is required to conclude this phase transition-like behavior as no anomaly is observed in resistivity and Seebeck coefficient, which could be sensitive to phase transition. Detailed structural/transport analysis at high-temperature will be investigated in our future work.

**Conclusions**

We investigated the synthesis and transport properties of As-doped layered pnictogen oxyselenides NdO$_{0.8}$F$_{0.2}$Sb$_{1-x}$As$_x$Se$_2$ ($x \leq 0.6$) for the first time. Despite the theoretical prediction of high-performance thermoelectric properties in these compounds, the experimentally obtained products were still insulators. We believe that local structure analysis will provide insights into the origin of the insulating behavior. Such studies are in progress and will be presented in our future work.

**Acknowledgement**

We thank fruitful discussion with K. Kuroki, M. Ochi, and H. Usui. This work was partly supported by JST CREST (No. JPMJCR16Q6), JSPS KAKENHI (No. 19K15291), and Advanced Research Program under the Human Resources Funds of Tokyo (H31-1).

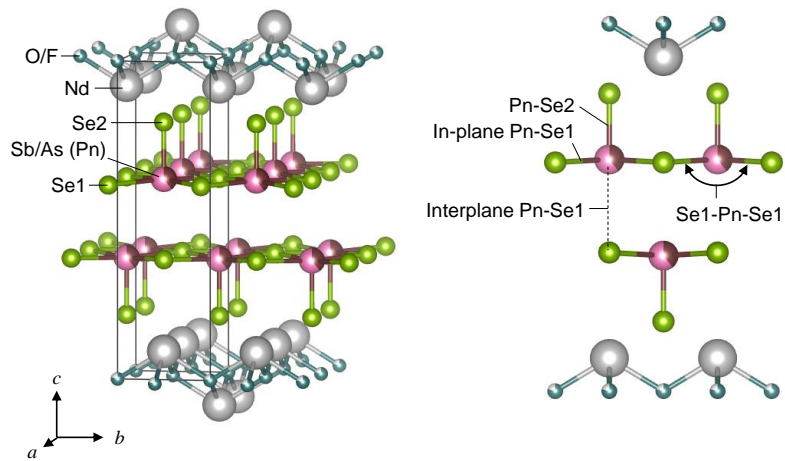

Figure 1

Schematic representation of the crystal structure of $NdO_{0.8}F_{0.2}Sb_{1-x}As_xSe_2$ ($x = 0.6$). The crystal structure belongs to tetragonal *P*4/*nmm* space group (No. 129). The unit cell is represented using the black line. Two crystallographic sites of Se are described as in-plane Se1 and out-of-plane Se2. Pn indicates pnictogen (Sb and As).

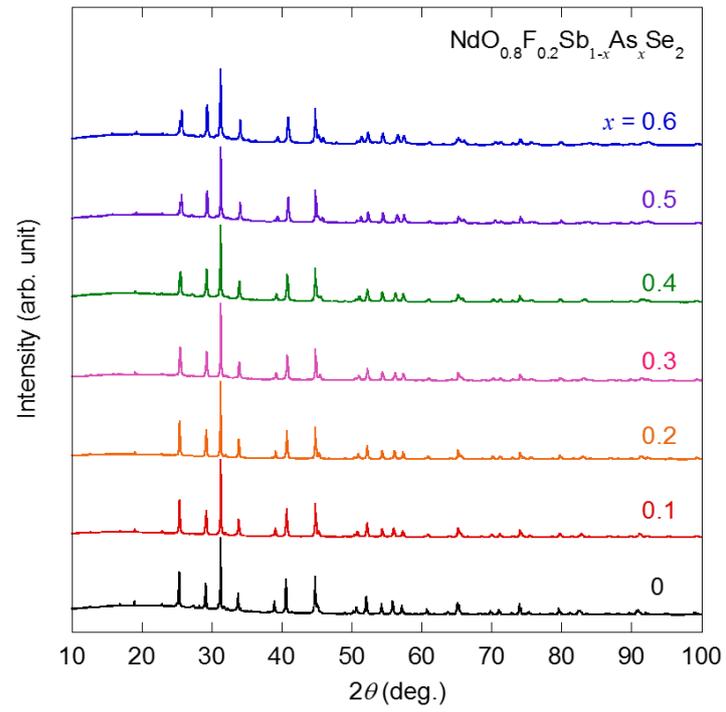

Figure 2

X-ray diffraction (XRD) patterns of NdO$_{0.8}$F$_{0.2}$Sb$_{1-x}$As$_x$Se$_2$ for $x$ = 0–0.6.

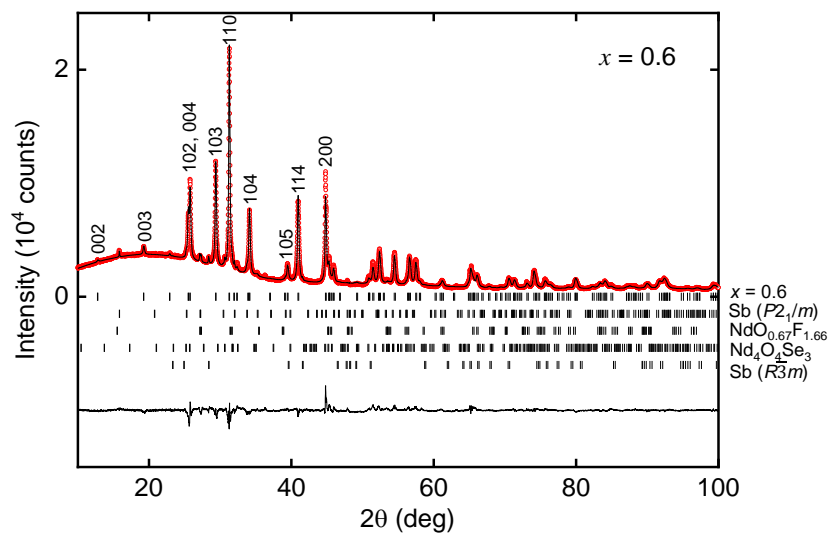

Figure 3

XRD pattern of $x = 0.6$ analyzed by the Rietveld method. The circles (red) and line (black) denote the observed and calculated patterns, respectively. The line at the bottom shows the difference between the observed and calculated intensities. The short bars represent the Bragg diffraction angles for $x = 0.6$ (93.4 wt.%) and impurities, Sb in the $P2_1/m$ space group (2.4 wt.%), NdO$_{0.67}$F$_{1.66}$ (2.0 wt.%), Nd$_4$O$_4$Se$_3$ (1.4 wt.%), and Sb in $R\bar{3}m$ space group (0.8 wt.%.), from top to bottom.

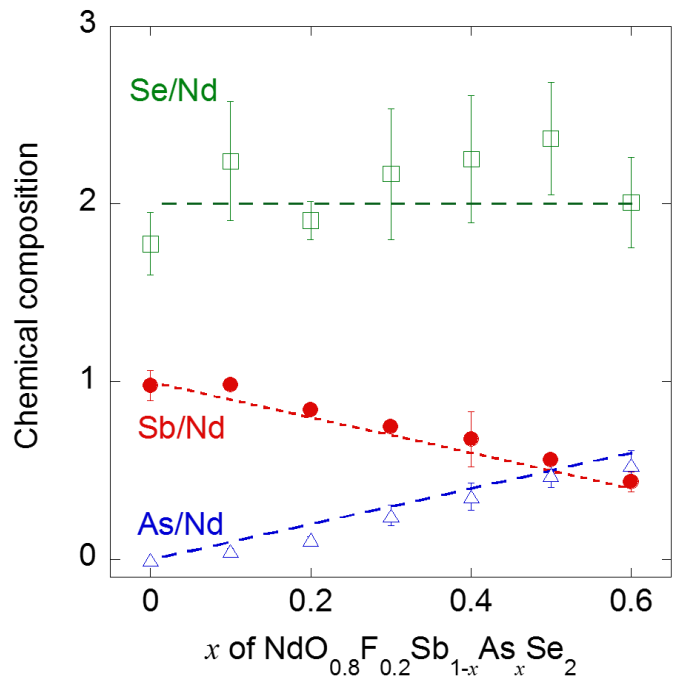

Figure 4

Chemical compositions obtained using EDX. The results are normalized with respect to Nd contents. The dashed lines represent the nominal compositions of the starting materials.

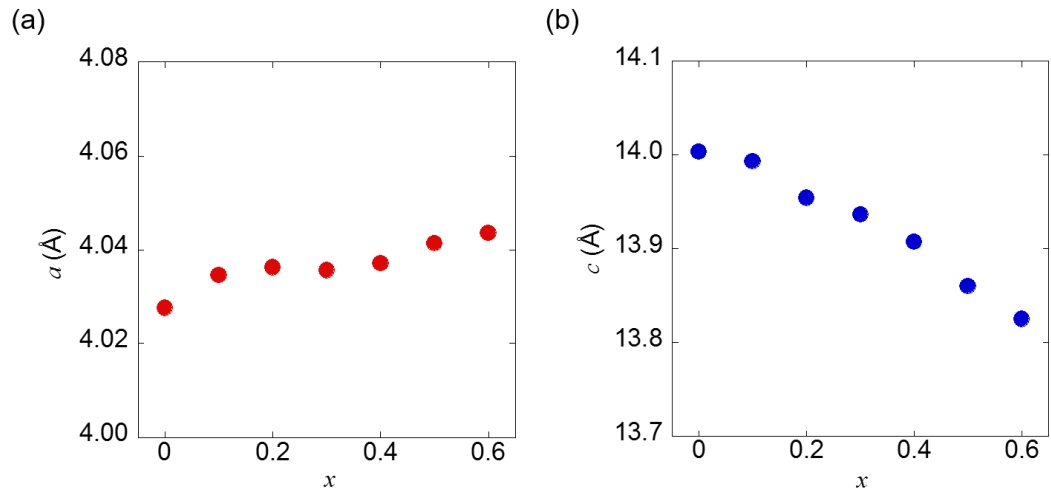

Figure 5

Lattice parameters (a) $a$ and (b) $c$ of NdO$_{0.8}$F$_{0.2}$Sb$_{1-x}$As$_x$Se$_2$.

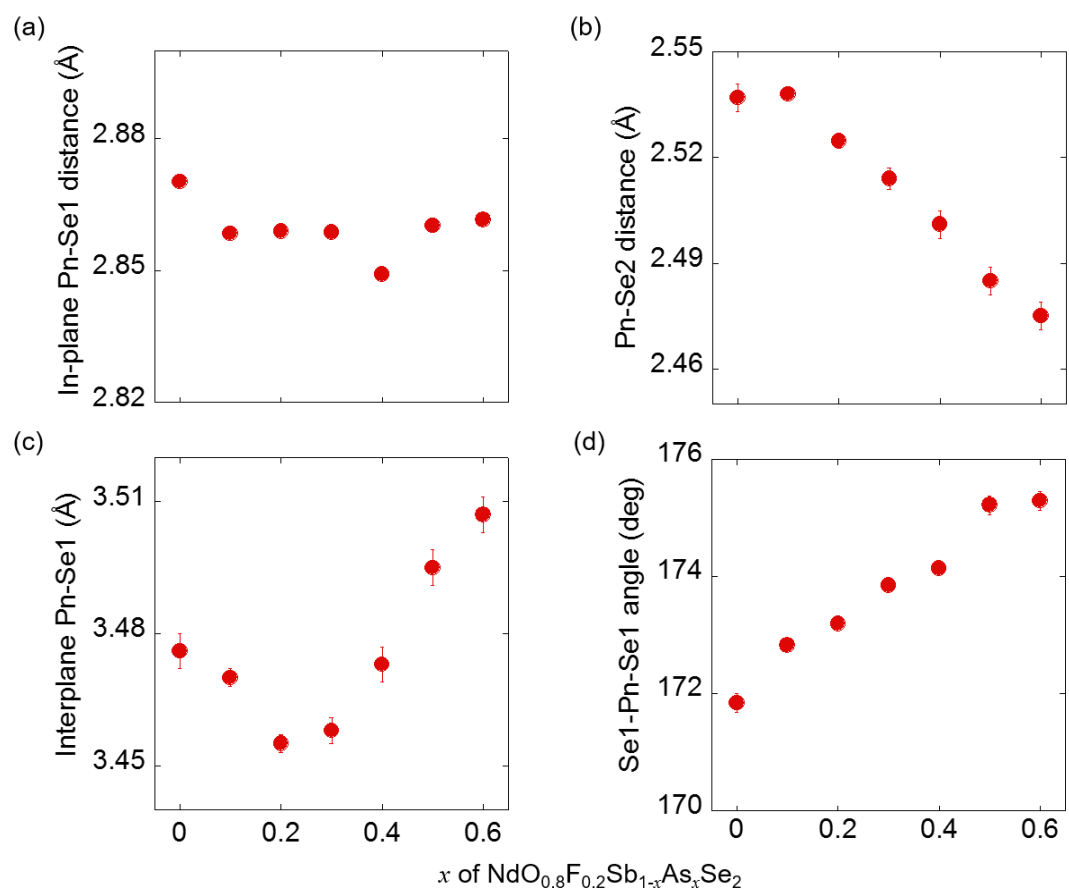

Figure 6

Selected bond distances and angle for $NdO_{0.8}F_{0.2}Sb_{1-x}As_xSe_2$: (a) in-plane Pn-Se1 distance, (b) Pn-Se2 distance, (c) interplane Pn-Se1 distance, and (d) Se1-Pn-Se1 angle.

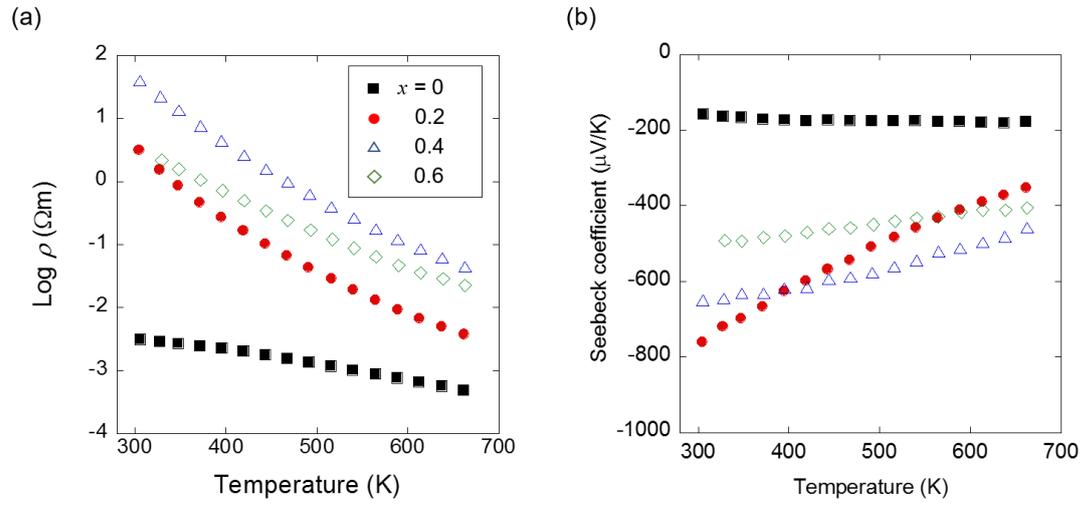

Figure 7

Temperature dependence of electrical transport properties of $NdO_{0.8}F_{0.2}Sb_{1-x}As_xSe_2$: (**a**) Electrical resistivity ($\rho$) and (**b**) Seebeck coefficient.

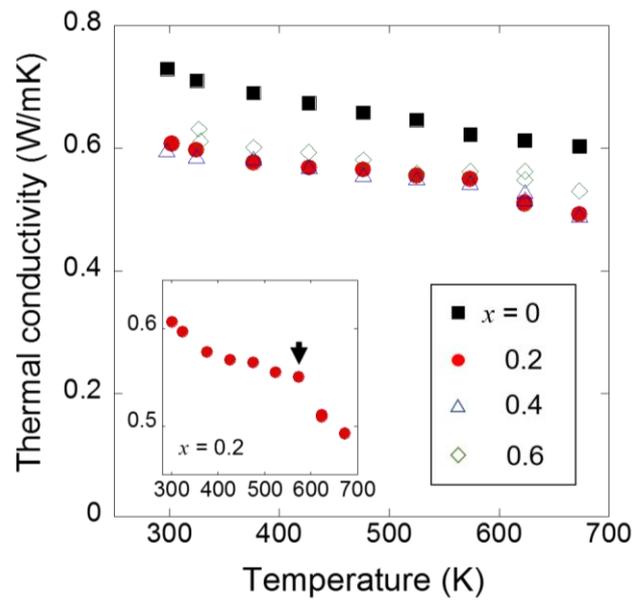

Figure 8

Temperature dependence of thermal conductivity for NdO$_{0.8}$F$_{0.2}$Sb$_{1-x}$As$_x$Se$_2$. The inset shows the expanded view of $x = 0.2$. The arrow indicates the temperature at which the thermal conductivity rapidly decreases.

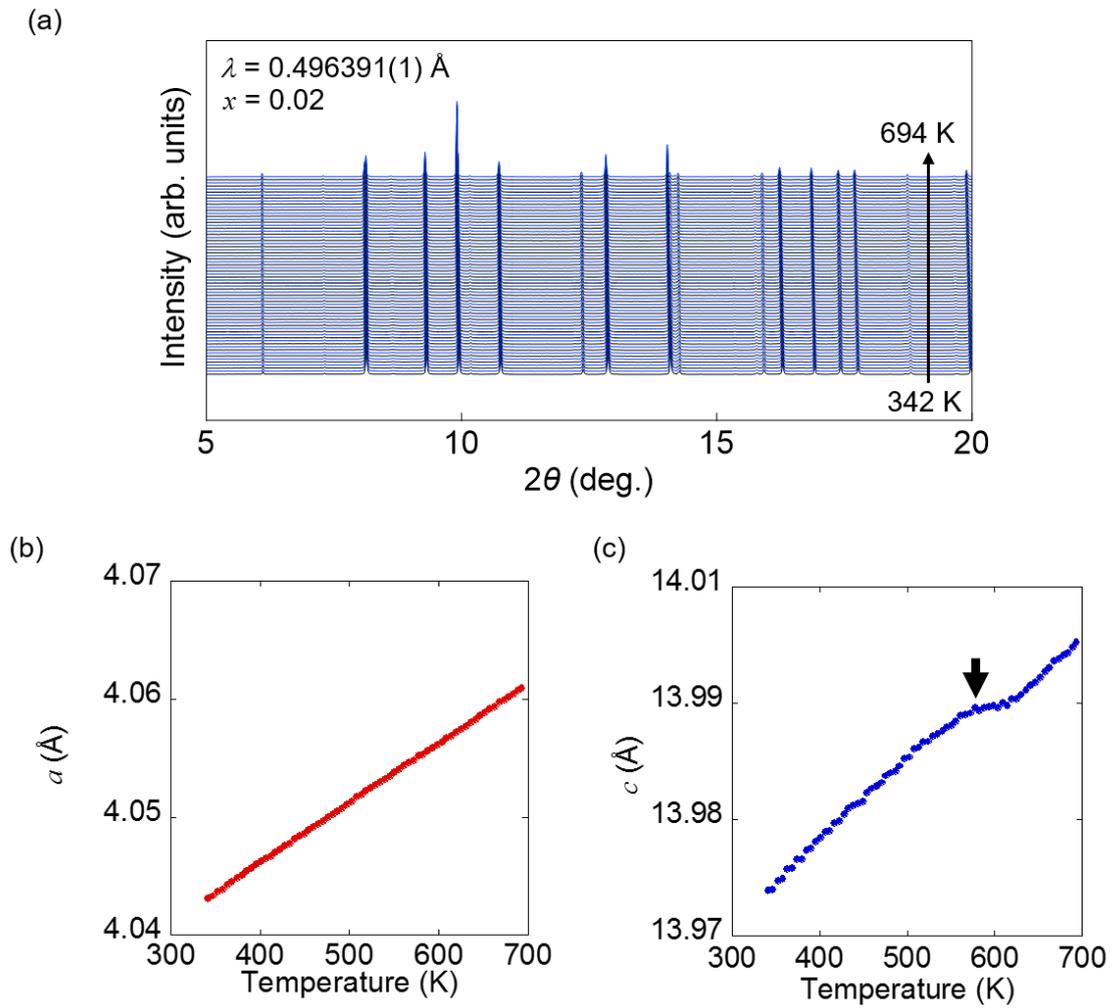

Figure 9

(a) Temperature dependence of synchrotron powder X-ray diffraction (SXRD) pattern of $x = 0.2$ at temperatures between 342 and 694 K. (b,c) Temperature dependence of lattice parameters (*a* and *c*) of $x = 0.2$. The arrow in (c) indicates the temperature at which an anomalous change in thermal conductivity is observed.